\documentclass[conference, a4paper]{IEEEtran}
\IEEEoverridecommandlockouts
\usepackage{cite}
\usepackage{amsmath,amssymb,amsfonts}
\usepackage{algorithmic}
\usepackage{graphicx}
\usepackage{textcomp}
\usepackage{xcolor}
\usepackage{bm}
\usepackage{caption}
\def\BibTeX{{\rm B\kern-.05em{\sc i\kern-.025em b}\kern-.08em
    T\kern-.1667em\lower.7ex\hbox{E}\kern-.125emX}}

\IEEEpubid{\makebox[\columnwidth]{978-1-5386-8235-7/18/\$31.00~\copyright~2018 IEEE \hfill} 
\hspace{\columnsep}\makebox[\columnwidth]{}}

\begin{document}

\title{FPGA Based SoC Estimator and Constant Current\\
		Charging/Discharging Controller for\\
			Lead--Acid Battery}

\author{P.~Dinesh,
		K.~Kumar~Teja,
        Shashank~Singh,
        Selvan~M.P.,
        and~Moorthi~S.
        \\\textit{Hybrid Electrical Systems Laboratory, Department of Electrical and Electronics Engineering}
        \\\textit{National Institute of Technology Tiruchirappalli, Tamil Nadu 620015, India}
        \\{\{p.dinesh.id, kumartejakalava, shashanksingh0110\}@gmail.com, \{selvanmp, srimoorthi\}@nitt.edu}
\thanks{Funded under the Visvesvaraya PhD Scheme of Ministry of Electronics and Information Technology, Government of India, being implemented by Digital India Corporation (formerly Media Lab Asia).}        
        }

\maketitle

\begin{abstract}
The state of charge (SoC) and the rate of charging/discharging current are the vital parameters associated with a battery by which its accurate runtime can be estimated. This paper aims to design a controller which comprises of a field--programmable gate array, back to back connected dc--dc converters, and a resistive touch display based graphical user interface (GUI). The controller estimates SoC and performs constant current charging or discharging of the battery. The implementation of GUI is to input the reference charging/discharging current from the user and to display the SoC. This research delves into battery and obtains its internal parameters by conducting hybrid pulse power characterization test. Furthermore, the obtained internal parameters are processed through an extended Kalman filter which yields the SoC. This controller has large applications in the renewable energy system, battery testing system, smart residential energy management systems, and micro--grids.
\end{abstract}
\vspace*{.5em}
\renewcommand\IEEEkeywordsname{Keywords}

\section{Introduction}
The effects of climate change and the extinction of fossil fuels have led to the rapid development of clean renewable energy generation systems to meet the ever-increasing demand for electricity. There is a need of energy storage system to meet the variations in demand. Batteries are known to be one of the best electrical energy storage systems. Among various batteries, the lead-acid batteries are preferred for many applications because of its low-cost, reliability, robustness, tolerant to overcharging, and can deliver very high constant currents. The extensive applications of lead-acid batteries are in electric vehicles, renewable energy integration, uninterruptible power supplies, emergency lights, etc. The lead-acid batteries are observed to be a cost-effective solution of an energy storage system with high power density and operational safety.

The state of charge (SoC) is a measure of the available amount of charge in the battery. The complex internal dynamics of batteries makes this parameter difficult to estimate. It is a critical parameter, which if kept within appropriate limits (e.g. 20\% to 80\%), not only improves the lifetime of battery by preventing it from being overcharged or deep discharged but also increases its reliability in various applications \cite{R1}, \cite{R2}. Different methods have been developed for the SoC estimation, e.g. specific gravity, open circuit voltage (OCV), and coulomb counting. In specific gravity method, the measurement of concentration of electrolyte is required for estimation of the SoC. It requires large stabilization time for getting satisfactory results. The OCV has a direct relationship with SoC which requires battery to be kept in rest position. The current integration method causes drift in results over long period because of error accumulation. 

To overcome these drawbacks, various equivalent circuit model (ECM) based algorithms are proposed, e.g. extended Kalman filter (EKF) \cite{R3,R4,R5,R6}, unscented Kalman filter \cite{R7,R8,R9}, and discrete non-linear observer \cite{R10}. These algorithms are more robust because it requires the measurement of current and voltage unlike current integration method or open circuit voltage method which relies only upon one parameter. Further, the measured terminal voltage acts as a feedback to form a closed-loop estimation method which in turn provides more accurate results. The advancement in machine learning and artificial intelligence in recent years have led to the development of neural network based algorithms \cite{R11}, \cite{R12}. The fuzzy model and support vector machine based algorithms are proposed in \cite{R13} and \cite{R14}. These methods \cite{R11, R12, R13, R14} require an intensive training with huge data and provide a powerful means of modeling a complex nonlinear system. Zhao H. \textit{et al.} \cite{R15} have proposed dual-polarization-resistance model and stated that the closed loop ECM based models are accurate and robust, and has found wide application for SoC estimation.

A key step towards SoC estimation is construction of ECM with high-fidelity. The ECM of the battery is obtained from hybrid pulse power characterization (HPPC) test. The HPPC test needs constant current charging/discharging of the battery. The applications of constant current charging/discharging are many, e.g. smart energy management systems \cite{R16, singh2018}, battery testing systems \cite{R17}, micro-grids \cite{R18}, electric vehicles \cite{R19}. The design and development of a controller using field-programmable gate array (FPGA) is the major contribution of this work. The controller's capability are as follows: (1) Performs HPPC test and obtains the values of internal parameters to construct the ECM. (2) Executes EKF to estimate SoC. (3) Charges/discharges battery at constant current rate. (4) Equipped with a touch-screen based graphical user interface (GUI) to input the reference value of charging/discharging current.

The paper is organized in following way. Section II deals with design and development of controller followed by constant current charging/discharging algorithm in section III. The battery modeling and literature are explained in section IV. Section V contains experiments performed to obtain internal parameters of battery, SoC measurements and results. Section VI draws the conclusion. 

\section{Description of Controller}
The developed controller charges/discharges the battery at a constant current provided by user via GUI. It also estimates the SoC of the battery. The controller comprises of two back to back connected DC-DC converters equipped with closed loop current control and a GUI. The GUI displays the SoC of the battery and provides interface to input the reference value of charging/discharging current.

The closed loop current control for constant charging/discharging heavily relies on accurate measurement of current. The accurate voltage measurement ensures operational safety of the battery and improves the accuracy of estimated SoC. The ACS-712 30A Hall effect current sensor module \cite{R20} and 5:1 voltage divider module are used as current sensing unit (CSU) and voltage sensing unit (VSU) respectively. The schematic of proposed controller is shown
in Fig.~\ref{fig1} below.

\begin{figure}[htbp]
\captionsetup{justification=raggedright,singlelinecheck=false}
\centerline{\includegraphics[width = \columnwidth]{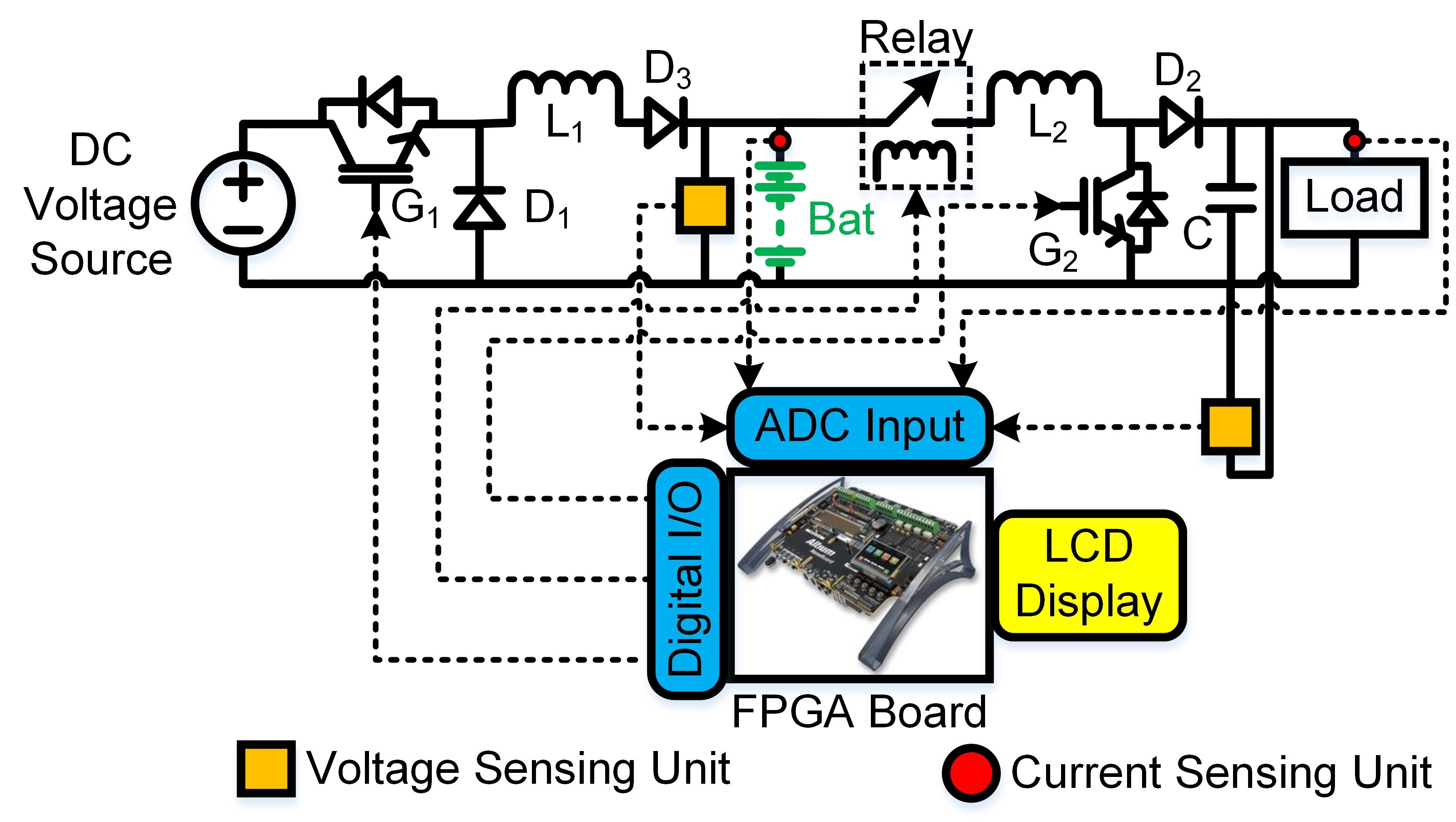}}
\caption{Schematic of proposed controller}
\label{fig1}
\end{figure}

\begin{figure}[htbp]
\captionsetup{justification=raggedright,singlelinecheck=false}
\centerline{\includegraphics[width = \columnwidth]{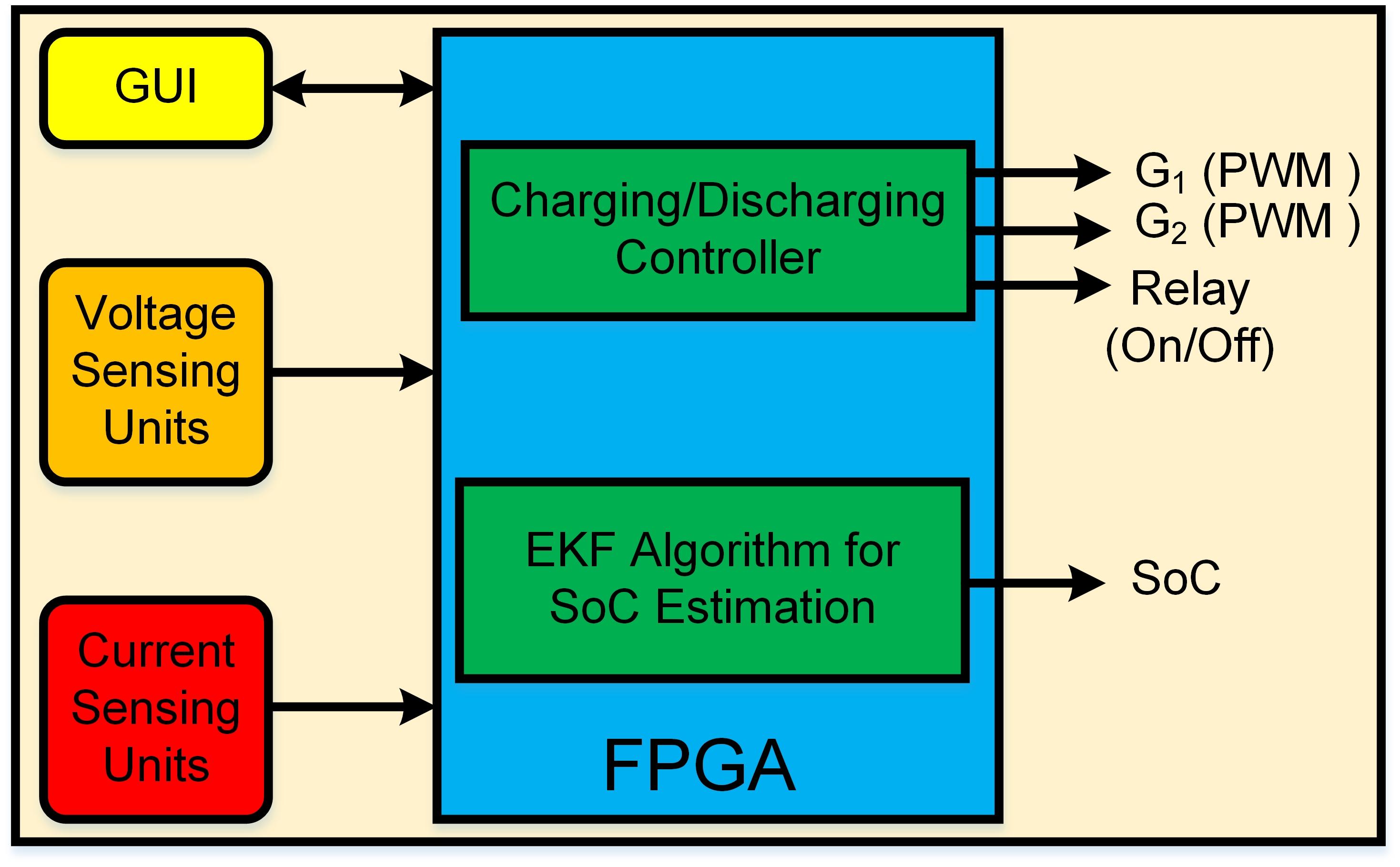}}
\caption{Block diagram of FPGA and its interfacing with peripherals}
\label{fig2}
\end{figure}

The modi operandi of controller are namely charging mode and discharging mode. A relay is used to switch between charging and discharging modes. An Altium Nanoboard 3000 with Xilinx SPARTAN-3AN FPGA is the processing unit which does following tasks (Fig.~\ref{fig2}): (1) Computation of voltage and current from signals acquired by analog to digital converter (ADC). (2) Generates square wave pulses for the operation of converters. (3) Computation of duty cycle of square wave pulses from user provided value of reference current to perform constant current charging or discharging. (4) Executes algorithm for estimation of SoC of the battery. (5) Trigger the relay contacts based on modes of operation.

\section{Constant Current Charging/Discharging}
\begin{figure}[htbp]
\captionsetup{justification=raggedright,singlelinecheck=false}
\centerline{\includegraphics[width = \columnwidth]{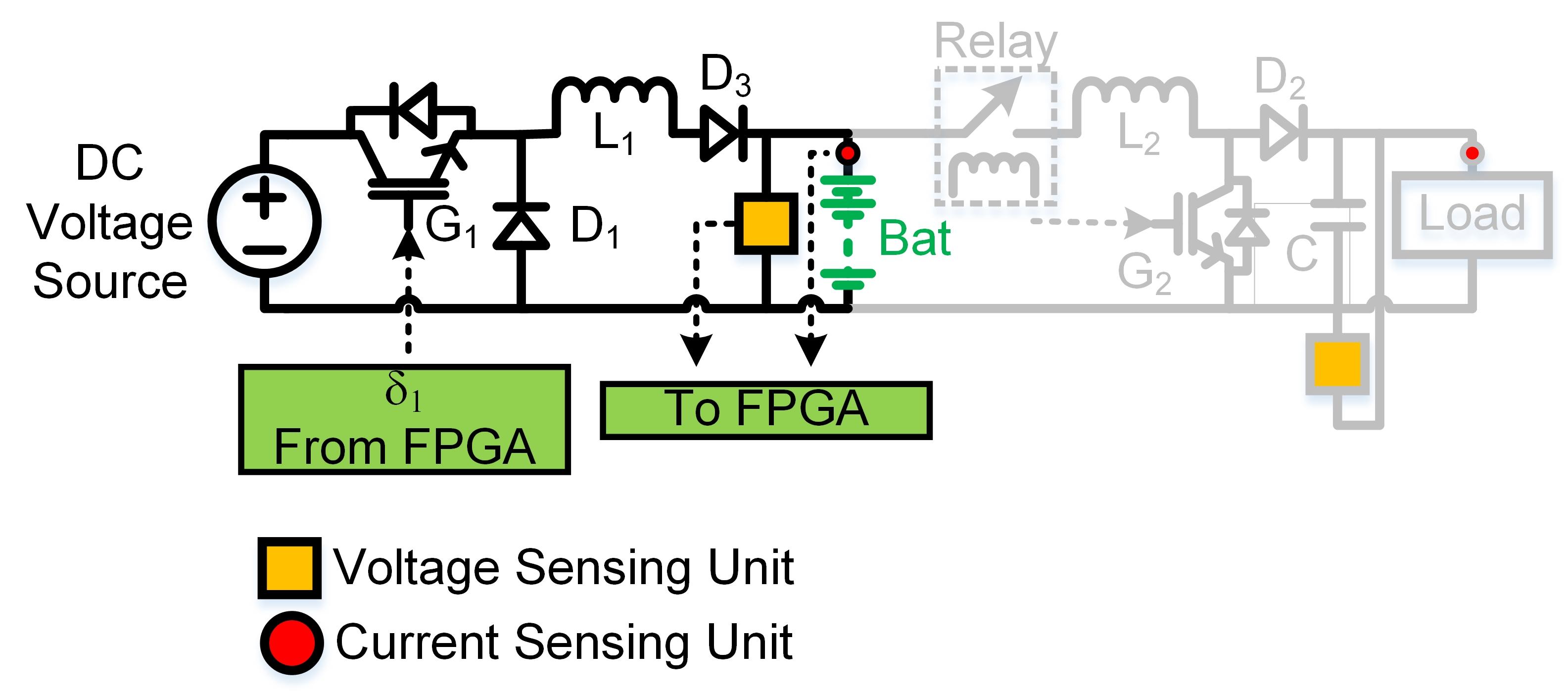}}
\caption{Charging mode of operation}
\label{fig3}
\end{figure}

\begin{figure}[htbp]
\captionsetup{justification=raggedright,singlelinecheck=false}
\centerline{\includegraphics[width = \columnwidth]{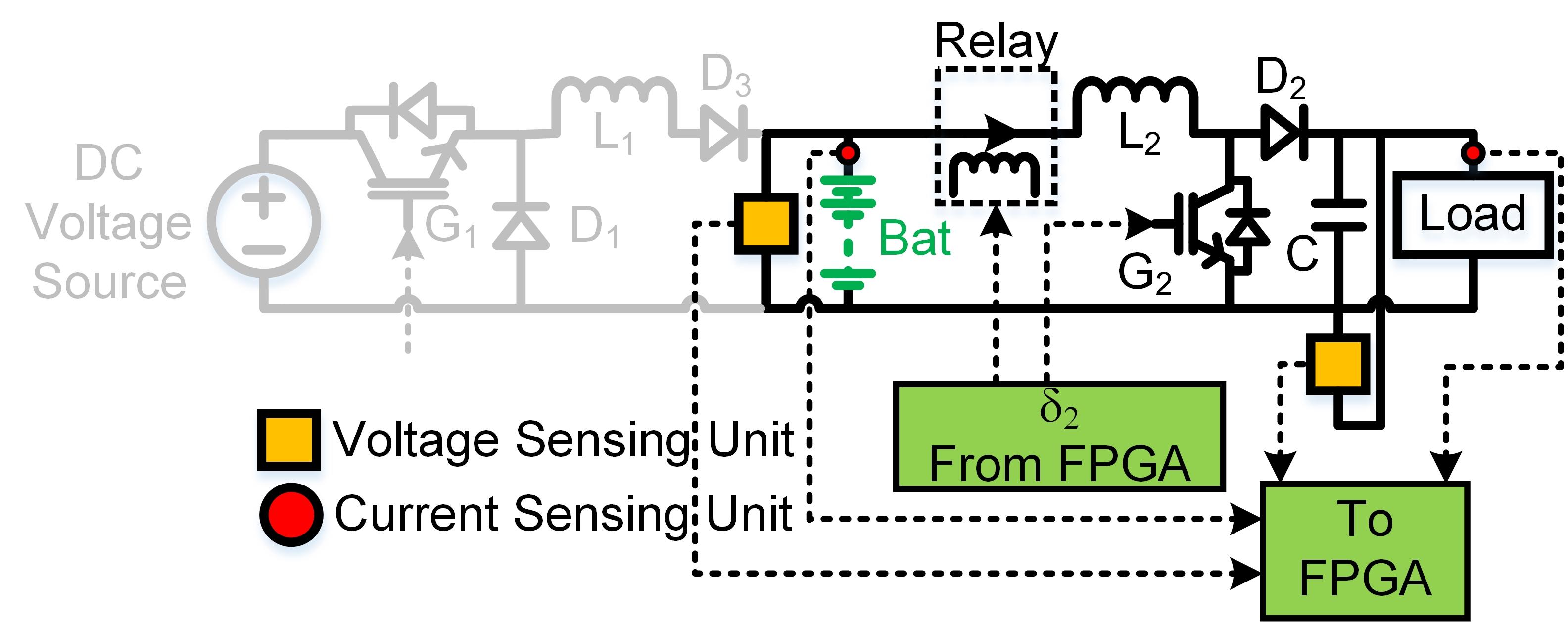}}
\caption{Discharging mode of operation}
\label{fig4}
\end{figure}

The charging/discharging modes of operation are shown in Fig.~\ref{fig3} and Fig.~\ref{fig4} respectively. It is evident from \eqref{e1} and \eqref{e2} that the setpoint value of charging/discharging current is a function of duty cycle. The algorithm implements a proportional-integral-derivative (PID) controller to estimate the required duty cycle.

\begin{equation}
I_{ref}^{charging}={{I}_{battery}}=\frac{{{I}_{source}}}{{{\delta }_{1}}}
\label{e1}
\end{equation}
where, $\delta_1$: duty cycle of pulse supplied to $G_1$
\begin{equation}
I_{ref}^{discharg ing}={{I}_{battery}}=\frac{{{I}_{load}}}{(1-{{\delta }_{2}})}
\label{e2}
\end{equation}
where,
$\delta_2$: duty cycle of pulse supplied to $G_2$\\

\begin{figure}[htbp]
\captionsetup{justification=raggedright,singlelinecheck=false}
\centerline{\includegraphics[width = \columnwidth]{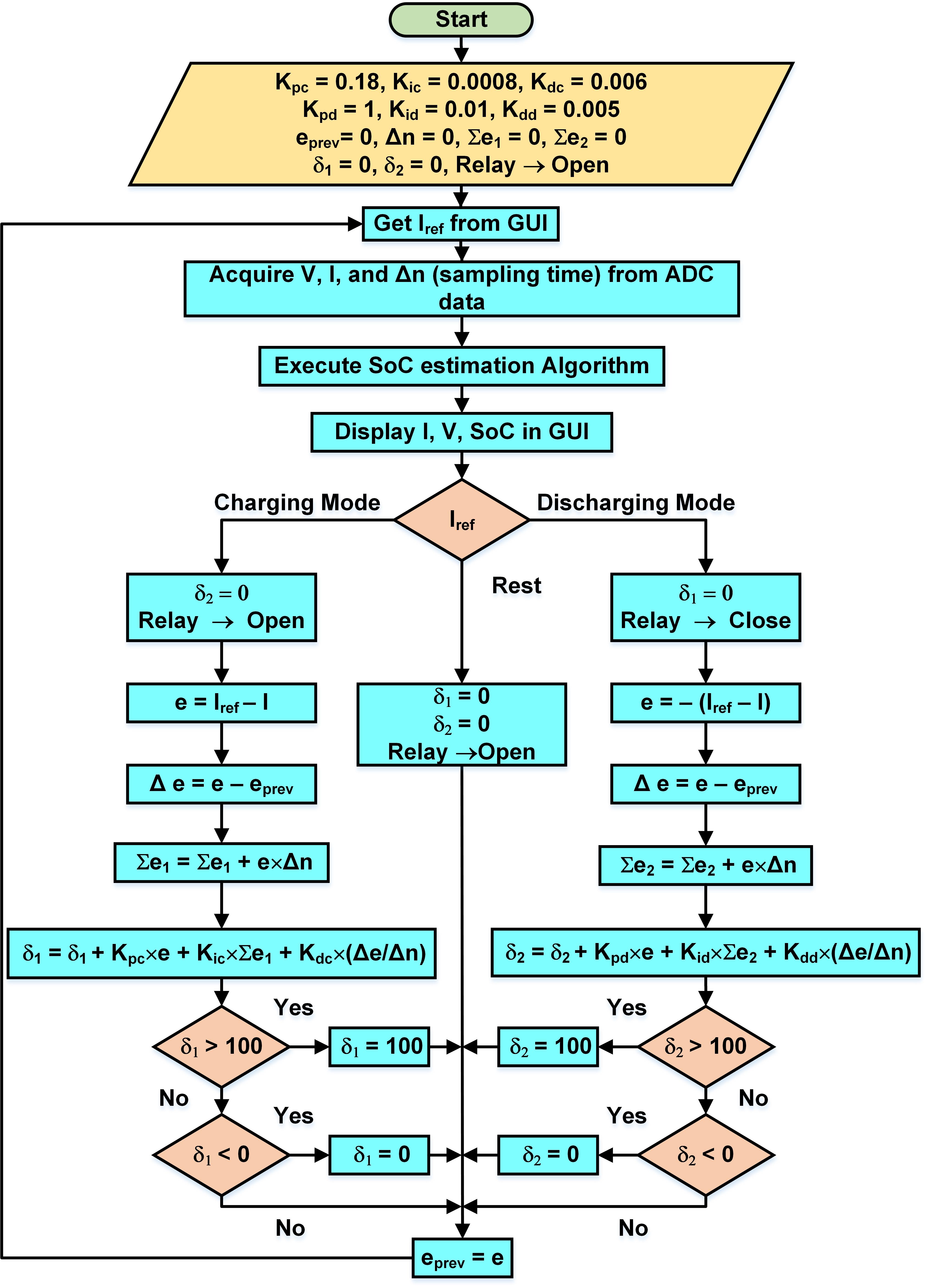}}
\caption{Flowchart of proposed algorithm}
\label{fig5}
\end{figure}

The setpoint value of charging/discharging current is obtained from user input. The instantaneous voltage and current data are acquired by VSU and CSU, and supplied to ADC of FPGA. Upon processing the ADC data, the error ($e$) between reference value and instantaneous value of currents is calculated, which is passed through a PID controller. The necessary duty cycle during charging/discharging mode is obtained using \eqref{e3}. The flowchart of the algorithm is shown in Fig.~\ref{fig5}.
\begin{equation}
\delta [n+1]=\delta [n]+{{K}_{p}}\times e[n]+{{K}_{i}}\times \sum{e[n]+{{K}_{d}}\times \frac{\Delta e}{\Delta n}}\
\label{e3}
\end{equation}
where, $\delta~(\delta_1, \delta_2)$ is duty cycle of converter $(G_1, G_2)$\\
$e[n]=I_{ref}^{charging/discharging}[n]-{{I}_{battery/load}}[n]$\\
$\Delta e=e-{{e}_{prev}}$\\
$Charging~Mode\Rightarrow [K_{pc}, K_{dc}, K_{ic}]=[0.18, 0.006, 0.0008]$\\
$Discharging~Mode\Rightarrow [K_{pd}, K_{dd}, K_{id}]=[1.0, 0.005, 0.01]$

\section{Battery Modeling}
\subsection{Development of equivalent circuit}
The SoC, $s(t)$ reflects status of available charge in the battery. The $s(t)$ is 100\% and 0\% for fully charged and discharged battery respectively. The total capacity $Q$ of a battery is defined as the aggregate amount of charge moved out of it during discharging from its $s(t) = 100\%~to~0\%$. The unit of $Q$ is ampere-hour (Ah) or milliampere-hour (mAh). The SoC can be modeled as \eqref{e4}, and is expressed in time-domain as \eqref{e5}, where $s(t_0)$ represents the amount of available charge at the instant, $t_0$.
\begin{equation}
\frac{\text{ds}}{\text{d}t}=\frac{i(t)}{Q}
\label{e4}
\end{equation}
\begin{equation}
s(t)=s({{t}_{0}})+\frac{1}{Q}\int_{{{t}_{0}}}^{t}{i}(\tau )d\tau
\label{e5}
\end{equation}
where, $i(t) < 0 \Rightarrow$ discharging current and $i(t) > 0 \Rightarrow$ charging current

Assuming constant current during sampling interval, the SoC can be expressed in discrete time-domain as \eqref{e6}. Since the batteries are not perfectly efficient, an efficiency factor is accommodated and represented in \eqref{e7}. This is known as coulombic (or charge) efficiency which represents the coulombs get out of the battery for every coulomb get into the battery. 
\begin{equation}
s[n+1]=s[n]+i[n]\frac{\Delta n}{Q}
\label{e6}
\end{equation}
\begin{equation}
s[n+1]=s[n]+i[n]\times \eta \times \frac{\Delta n}{Q}
\label{e7}
\end{equation}
where,
$s[n]$: SoC at nth sampling interval\\
$\Delta n$: sampling interval\\
$\eta$: coulombic efficiency  

\begin{figure}[htbp]
\captionsetup{justification=raggedright,singlelinecheck=false}
\centerline{\includegraphics[width = 0.6\columnwidth]{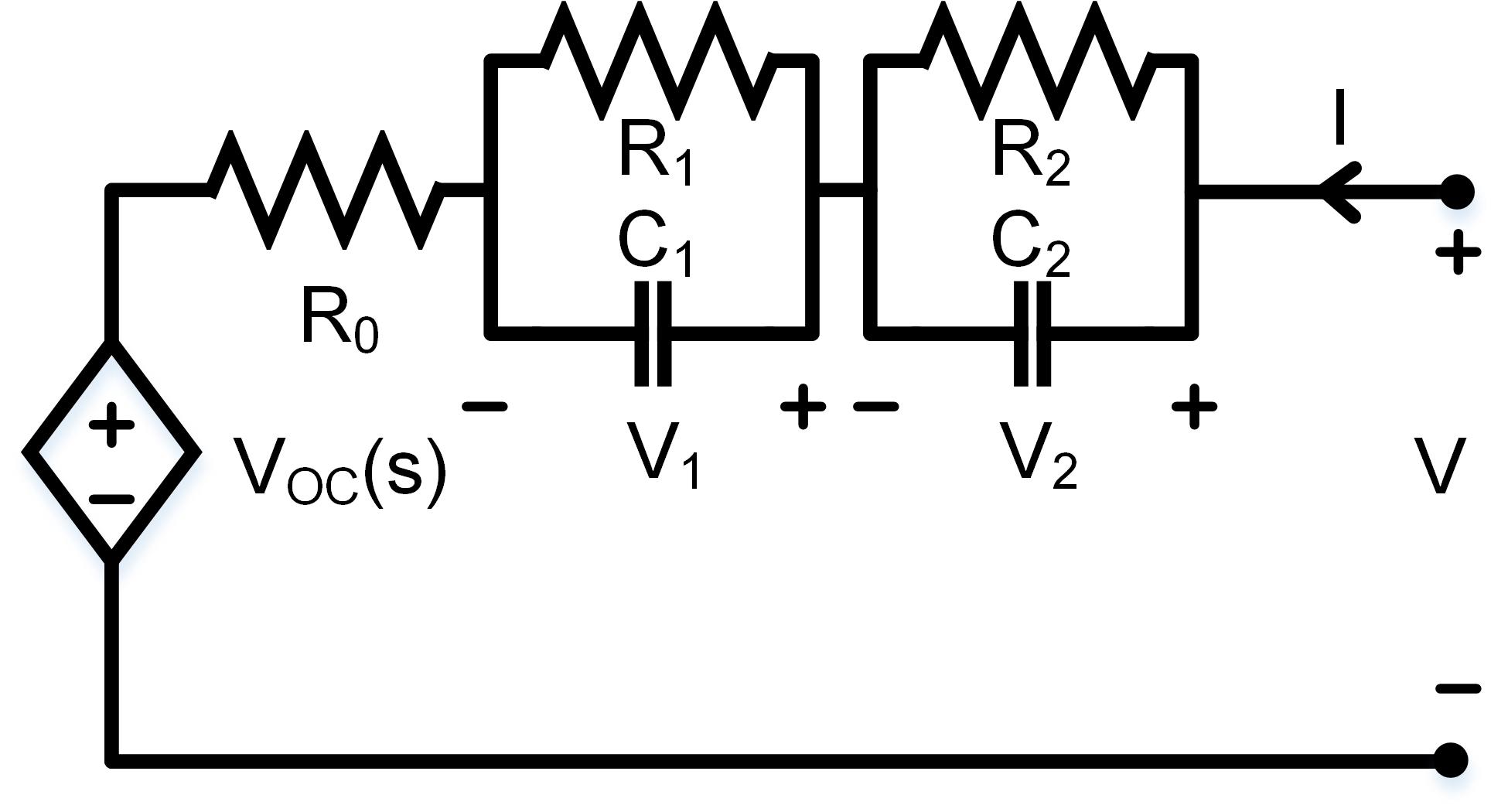}}
\caption{ECM of battery}
\label{fig6}
\end{figure}

Presently, there are no sensors available to measure the SoC, therefore it is estimated by measuring terminal voltage and battery current. The battery's open circuit voltage (OCV) is a function of SoC, hence it can be modeled as dependent voltage source. The battery is modeled by phenomenological analogs using circuit parameters. Considering the dynamic characteristics of a real battery, an ECM of the battery is depicted in Fig.~\ref{fig6}. The ECM comprises of a dependent voltage source, an equivalent ohmic resistance and infinite parallel RC pairs to represent the nonlinear polarization characteristics. The high number of the parallel RC pairs increases the complexity of the model. To overcome that, a $2^{nd}$ order ECM (2 RC pairs) is adapted in this research which is reasonable in complexity.

where,\\
$V_{OC}$: open circuit voltage\\ 
$V$: terminal voltage\\
$I$: current flowing into the battery\\
$R_0$: ohmic resistance\\
$R_1,~C_1,~R_2,~C_2$: represents non-linearity\\

From Fig.~\ref{fig6}, the electrical behavior model can be described by \eqref{e8}, \eqref{e9}, and \eqref{e10}.
\begin{equation}
\frac{d{{V}_{1}}}{dt}=-{{\frac{{{V}_{1}}}{{{R}_{1}}C}}_{1}}+\frac{I}{{{C}_{1}}}
\label{e8}
\end{equation}
\begin{equation}
\frac{d{{V}_{2}}}{dt}=-{{\frac{{{V}_{2}}}{{{R}_{2}}C}}_{2}}+\frac{I}{{{C}_{2}}}
\label{e9}
\end{equation}
\begin{equation}
V={{V}_{oc}}+I{{R}_{0}}+I{{R}_{1}}(1-{{e}^{\frac{-t}{{{R}_{1}}{{C}_{1}}}}})+I{{R}_{2}}(1-{{e}^{\frac{-t}{{{R}_{2}}{{C}_{2}}}}})
\label{e10}
\end{equation}

\section{Experimental Study}
\subsection{Estimation of Internal Parameters of Battery}
\begin{figure}[htbp]
\captionsetup{justification=raggedright,singlelinecheck=false}
\centerline{\includegraphics[width = 0.7\columnwidth]{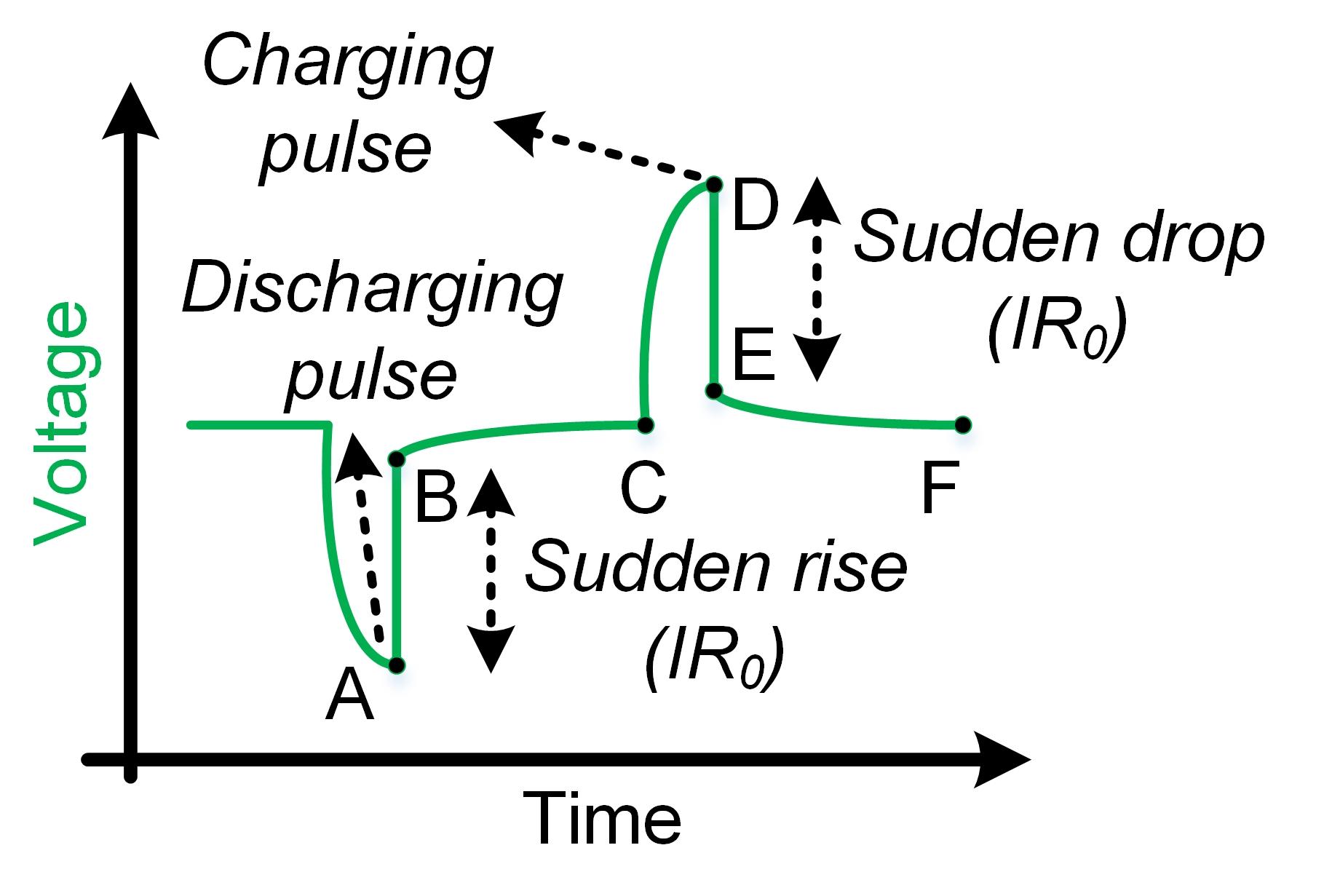}}
\caption{Voltage profile of battery during HPPC test}
\label{fig7}
\end{figure}

The parameters of the ECM depend on the SoC and the direction of current accounting for hysteresis characteristics. To determine these parameters at a certain SoC, two pulses namely an \textit{impulse charge} and an \textit{impulse discharge} should be applied to a battery at rest. The voltage profile for this test follows the curve shown in Fig.~\ref{fig7}.

From the curve (Fig.~\ref{fig7}), impulse rise and drop in voltage is evident. The illustrated rise ($AB$ section of curve) and drop ($DE$ section of curve) are the result of an \textit{impulse discharge} and an \textit{impulse charge} pulse respectively, and represent voltage drop across the ohmic resistor ($IR_0$). The sections $BC$ and $EF$ represent rest period. The remaining portion of curve ($BC$ and $EF$ section of curve) should be fitted into the \eqref{e11} using MATLAB curve fitting tool to obtain the parameters representing non-linearity in the battery. The internal ohmic resistance can be obtained using \eqref{e12}.
\begin{equation}
\frac{V-{{V}_{oc}}-I{{R}_{0}}}{I}=\alpha (1-{{e}^{\frac{-t}{\beta }}})+\gamma (1-{{e}^{\frac{-t}{\lambda }}})
\label{e11}
\end{equation}
\begin{equation}
{{R}_{0}}=\left| \frac{{{V}_{oc}}-V}{I} \right|
\label{e12}
\end{equation}
From the best fit, the parameters will be as follows:\\
$R_1=\alpha$, $R_2=\gamma$, $C_1=\frac{\beta}{\alpha}$, $C_2=\frac{\gamma}{\lambda}$

\begin{figure}[htbp]
\captionsetup{justification=raggedright,singlelinecheck=false}
\centerline{\includegraphics[width = \columnwidth]{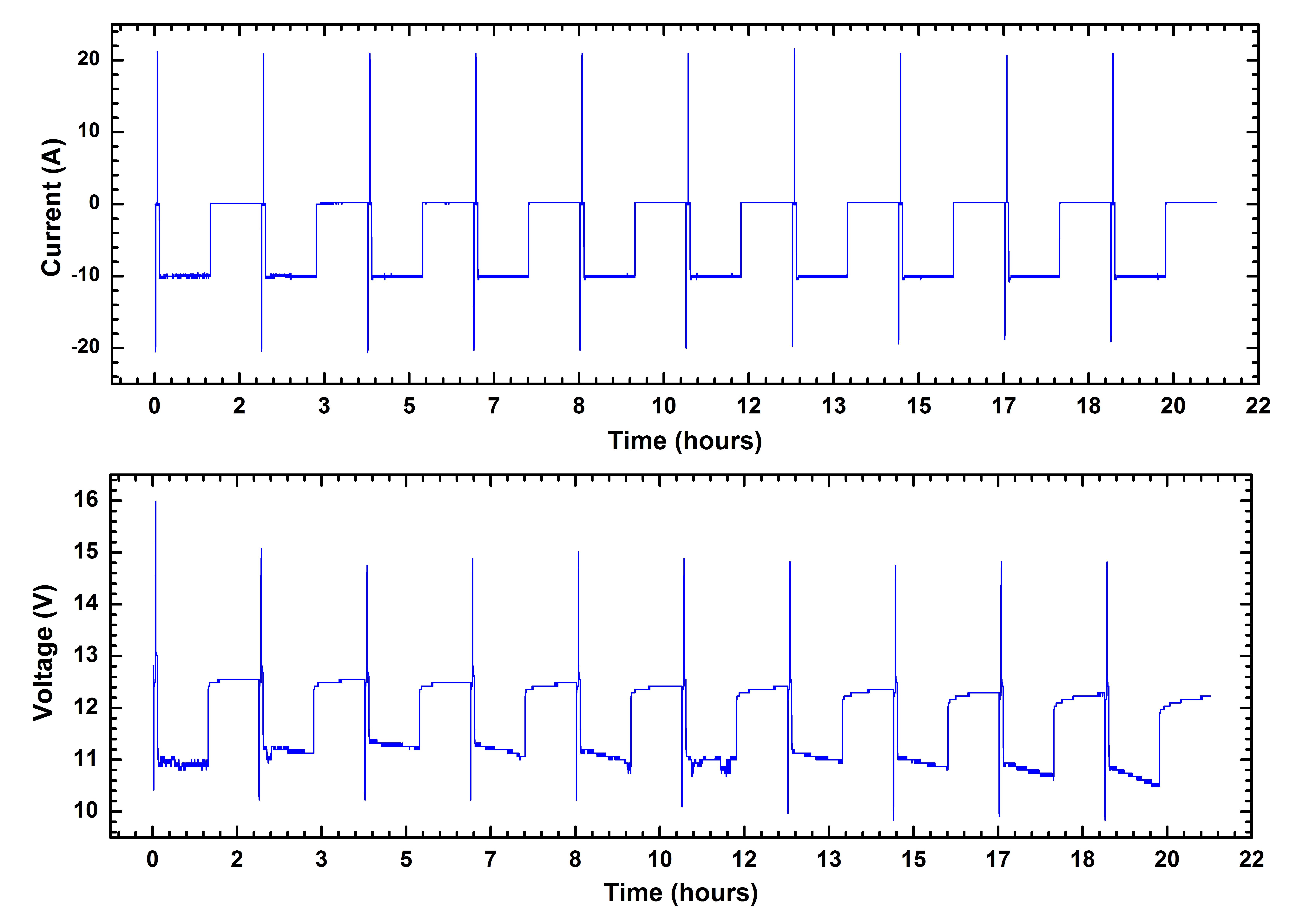}}
\caption{Current and Voltage test profiles}
\label{fig8}
\end{figure}

The similar test must be performed at every $10\%$ intervals of SoC starting from $s(t) = 100\%$. From the experiments, the obtained current test profile and voltage test profile of battery are depicted in Fig.~\ref{fig8}.
\begin{figure}[htbp]
\captionsetup{justification=raggedright,singlelinecheck=false}
\centerline{\includegraphics[width = \columnwidth]{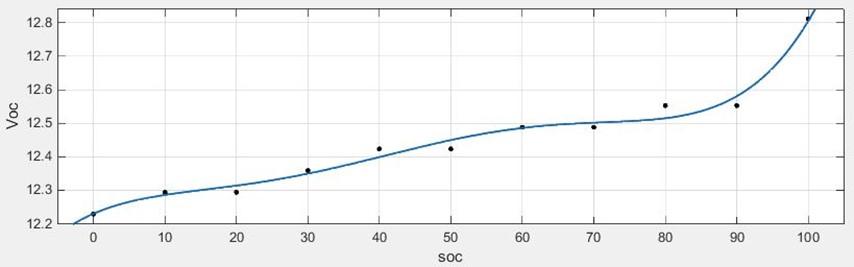}}
\caption{SoC versus OCV curve}
\label{fig9}
\end{figure}
Upon analyzing results from HPPC test, obtained parameters during charging/discharging at different SoC are tabulated in Table I and Table II respectively. Eventually, the relation between SoC and $V_{oc}$ is essential for SoC estimation, therefore data depicting variations in SoC are recorded from HPPC test during charging. The corresponding $V_{oc}$ values are taken at every $10\%$ intervals of SoC after giving one-hour rest period to the battery.  A $5^{th}$ degree polynomial is employed to fit the variation using MATLAB. The fitted curve is shown in Fig.~\ref{fig9} and expressed by \eqref{e13}.
\begin{table}[htbp]
\caption{Charging Data}
\begin{center}
\setlength\tabcolsep{4pt} 
\begin{tabular}{c|c|c|c|c|c}
\hline
\textbf{SoC} & \bm{$R_0~(m\Omega)$} & \bm{$R_1~(m\Omega)$} & \bm{$C_1~(kF)$} & \bm{$R_2~(m\Omega)$} & \bm{$C_2~(kF)$} \\
\hline
0	&112.318	&16.21	&0.937	&20.15	&0.750	\\	 
10	&110.695	&14.52	&6.873	&28.99	&0.671\\
20	&107.181	&17.82	&1.565	&18.93	&1.467\\
30	&103.883	&18.45	&0.875	&26.01	&3.236\\
40	&103.883	&38.90	&0.897	&20.74	&2.190\\
50	&105.506	&41.95	&0.792	&44.16	&0.616\\
60	&107.289	&17.86	&1.260	&21.12	&1.051\\
70	&097.865	    &45.99	&1.469	&46.31	&1.456\\
80	&093.830	    &19.43	&1.722	&17.45	&2.005\\
90	&105.511	&09.20	&0.780	&13.15	&0.546\\
100	&117.297	&10.57	&0.736	&46.75	&0.680\\
\hline
\end{tabular}
\label{tab1}
\end{center}
\end{table}
\begin{table}[htbp]
\caption{Discharging Data}
\begin{center}
\setlength\tabcolsep{4pt} 
\begin{tabular}{c|c|c|c|c|c}
\hline
\textbf{SoC} & \bm{$R_0~(m\Omega)$} & \bm{$R_1~(m\Omega)$} & \bm{$C_1~(kF)$} & \bm{$R_2~(m\Omega)$} & \bm{$C_2~(kF)$} \\
\hline
0	&118.152	&23.49	&0.447601	&2.328	&2.306946\\
10	&116.176	&09.81	&0.377278	&1.417	&1.430313\\ 
20	&116.176	&14.19	&0.340939	&1.982	&1.886134\\
30	&110.702	&10.84	&0.388916	&5.174	&0.772787\\
40	&114.272	&06.23	&0.319658	&3.128	&0.677029\\
50	&112.429	&03.51	&0.838898	&3.371	&0.932269\\
60	&105.512	&05.64	&0.514280	&3.923	&0.753494\\
70	&107.239	&04.64	&0.854056	&1.184	&2.098372\\
80	&105.615	&06.16	&0.503514	&1.885	&2.638010\\
90	&105.512	&06.33	&0.434482	&1.418	&1.766140\\
100	&099.014	&17.08	&0.417306	&4.515	&0.965072\\
\hline
\end{tabular}
\label{tab2}
\end{center}
\end{table}
\begin{equation}
{{V}_{oc}}={{\Lambda }_{1}}{{(S)}^{5}}+{{\Lambda }_{2}}{{(S)}^{4}}+{{\Lambda }_{3}}{{(S)}^{3}}+{{\Lambda }_{4}}{{(S)}^{2}}+{{\Lambda }_{5}}(S)+{{\Lambda }_{6}}
\label{e13}
\end{equation}
where, $\Lambda_1 = 11.41,~\Lambda_2 = ‒24.38,~\Lambda_3 = 17.85$\\
$\Lambda_4 = ‒5.233,~\Lambda_5 = 0.928,~\Lambda_6 = 12.33$
\subsection{SoC Estimation Algorithm}
Kalman filtering is an established technology for dynamic system state estimation in various fields like global positioning, target tracking, and navigation. It is often used to optimally estimate the internal states of a system in the presence of uncertain and indirect measurements. In recent years, it is being used to determine the SoC of the battery. Due to the nonlinear characteristic of batteries, EKF is a suitable tool which approximates nonlinear system to linear time-varying system. The state-space representation of battery model is must to implement EKF. The state-space model is formulated and expressed using \eqref{e14} and \eqref{e15}.\\
Let us consider, state vector, $x=[s,~V_1,~V_2]^T$\\
Output vector, $y = V$ and input vector, $u = I$ then,
\begin{equation}
\frac{dx}{dt}=\left[ \begin{matrix}
   0 & 0 & 0  \\
   0 & -\frac{1}{{{R}_{1}}{{C}_{1}}} & 0  \\
   0 & 0 & -\frac{1}{{{R}_{2}}{{C}_{2}}}  \\
\end{matrix} \right]x+\left[ \begin{matrix}
   Q^{-1}  \\
   C_{1}^{-1}\\
   C_{2}^{-1}\\
\end{matrix} \right]I
\label{e14}
\end{equation}
\begin{equation}
y={{V}_{1}}+{{V}_{2}}+{{V}_{oc}}+I{{R}_{0}}
\label{e15}
\end{equation}
To realize \eqref{e14}, \eqref{e15} in FPGA, a discrete state-space model is needed which is expressed in \eqref{e16} and \eqref{e17}.  
\begin{equation}
x[n+1]=A_n \times x[n]+B_n \times u[n]
\label{e16}
\end{equation}
\begin{equation}
y[n+1]=C_n \times x[n]+ D_n \times u[n]
\label{e17}
\end{equation}
where,
${{A}_{n}}=\left[ \begin{matrix}
   1 & 0 & 0  \\
   0 & {{e}^{\frac{-\Delta n}{{{R}_{1}}{{C}_{1}}}}} & 0  \\
   0 & 0 & {{e}^{\frac{-\Delta n}{{{R}_{2}}{{C}_{2}}}}}  \\
\end{matrix} \right]$\\
${{B}_{n}}=\left[ \begin{matrix}
   \frac{\eta \times \Delta n}{Q}  \\
   -{{R}_{1}}(1-{{e}^{\frac{-\Delta n}{{{R}_{1}}{{C}_{1}}}}})  \\
   -{{R}_{2}}(1-{{e}^{\frac{-\Delta n}{{{R}_{2}}{{C}_{2}}}}})  \\
\end{matrix} \right]$,
${{C}_{n}}=\left[ \begin{matrix}
   \frac{d{{V}_{oc}}}{ds} & 1 & 1  \\
\end{matrix} \right]$, $D_n=[R_0]$

\subsection{EKF Implementation}
EKF is a recursive algorithm which executes the six steps (shown in \textbf{Algorithm}), and converges to the true state. Initialization of this algorithm requires initial states with known uncertainty. 

\begin{algorithmic}[1]
\centerline{\textbf{Algorithm}}
\STATE\textbf{Step 1: \textit{State estimate time update}}
\STATE ${{\hat{x}}^{-}}[n+1]={{A}_{n}}\times \hat{x}[n]+{{B}_{n}}\times u[n]$
\STATE\textbf{Step 2: \textit{Error covariance time update}}
\STATE ${{P}^{-}}[n+1]={{A}_{n}}\times P[n]\times A_{n}^{T}+J$
\STATE\textbf{Step 3: \textit{Formulation of Kalman gain matrix}}
\STATE $K[n+1]=P^{-1}[n+1]C_{n+1}^T[C_{n+1}P^{-1}[n+1]C_{n+1}^T+R]^{-1}$
\STATE\textbf{Step 4: \textit{Predicted output}}
\STATE ${{\hat{y}}^{-}}[n+1]={{C}_{n+1}}\times {{\hat{x}}^{-}}[n+1]+{{D}_{n+1}}\times I[n+1]$
\STATE\textbf{Step 5: \textit{State estimate measured update}}
\STATE ${{\hat{x}}^{+}}[n+1]={{\hat{x}}^{-}}[n+1]+K[n+1]\left[ U[n+1]-{{{\hat{y}}}^{-}}[n+1] \right]$
\STATE\textbf{Step 6: \textit{Error covariance measurement update}}
\STATE ${{P}^{+}}[n+1]=\left[ 1-K[n+1]\times {{C}_{n+1}} \right]\times {{P}^{-}}[n+1]$
\end{algorithmic} 

Where, $R$ and $J$ are noise co-variance matrices of measurement and process respectively. $P$ is a co-variance matrix of state vector. ${{\hat{x}}^{-}}$ and $P^-$ are predicted prior estimates. ${{\hat{x}}^{+}}$ and $P^+$ are corrected posterior estimates.

\section{Results}
\begin{figure}[htbp]
\captionsetup{justification=raggedright,singlelinecheck=false}
\centerline{\includegraphics[scale = 0.055]{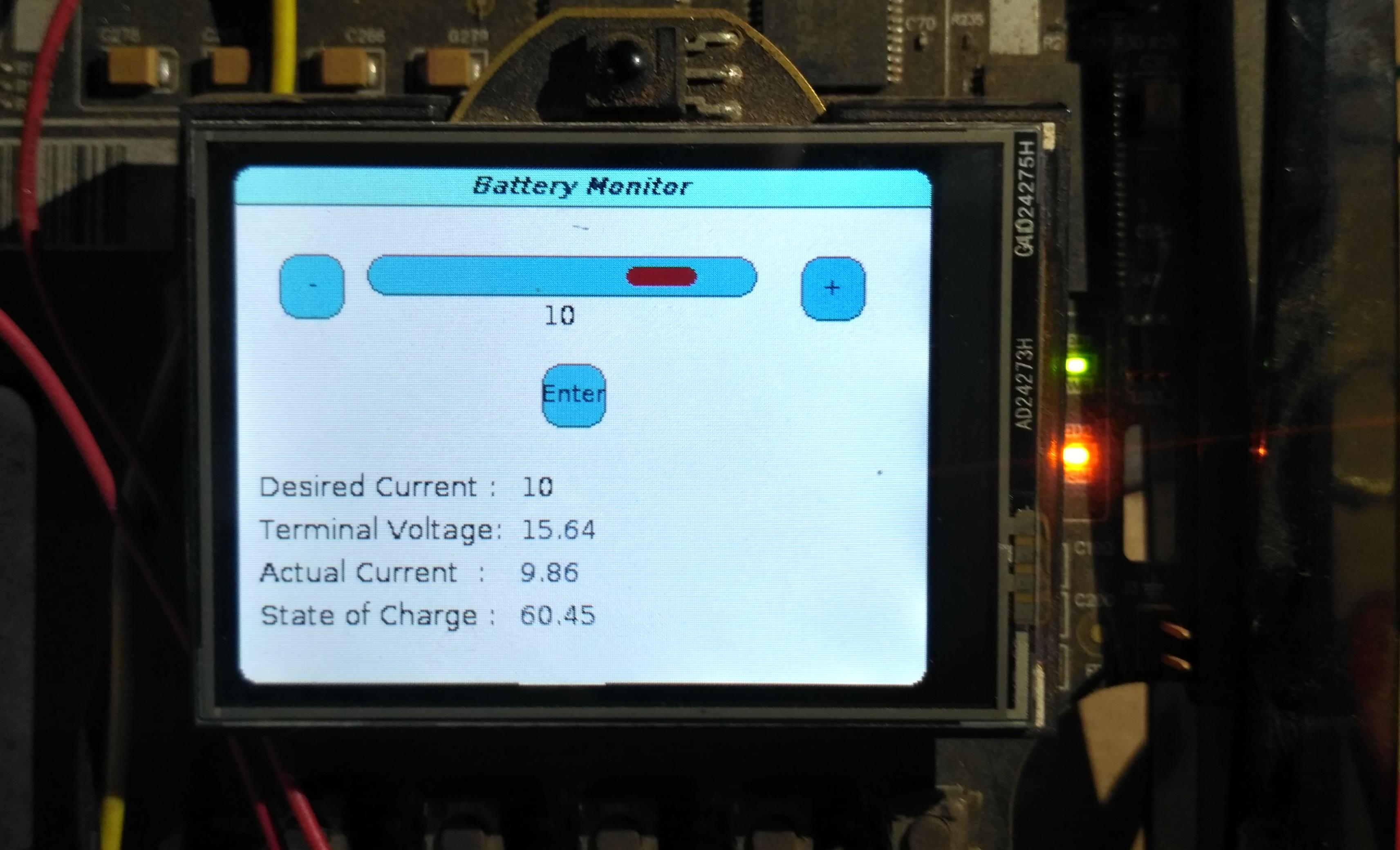}}
\caption{A sample image of developed GUI}
\label{fig10}
\end{figure}

\begin{figure}[t]
\captionsetup{justification=raggedright,singlelinecheck=false}
\centerline{\includegraphics[width = 0.8\columnwidth]{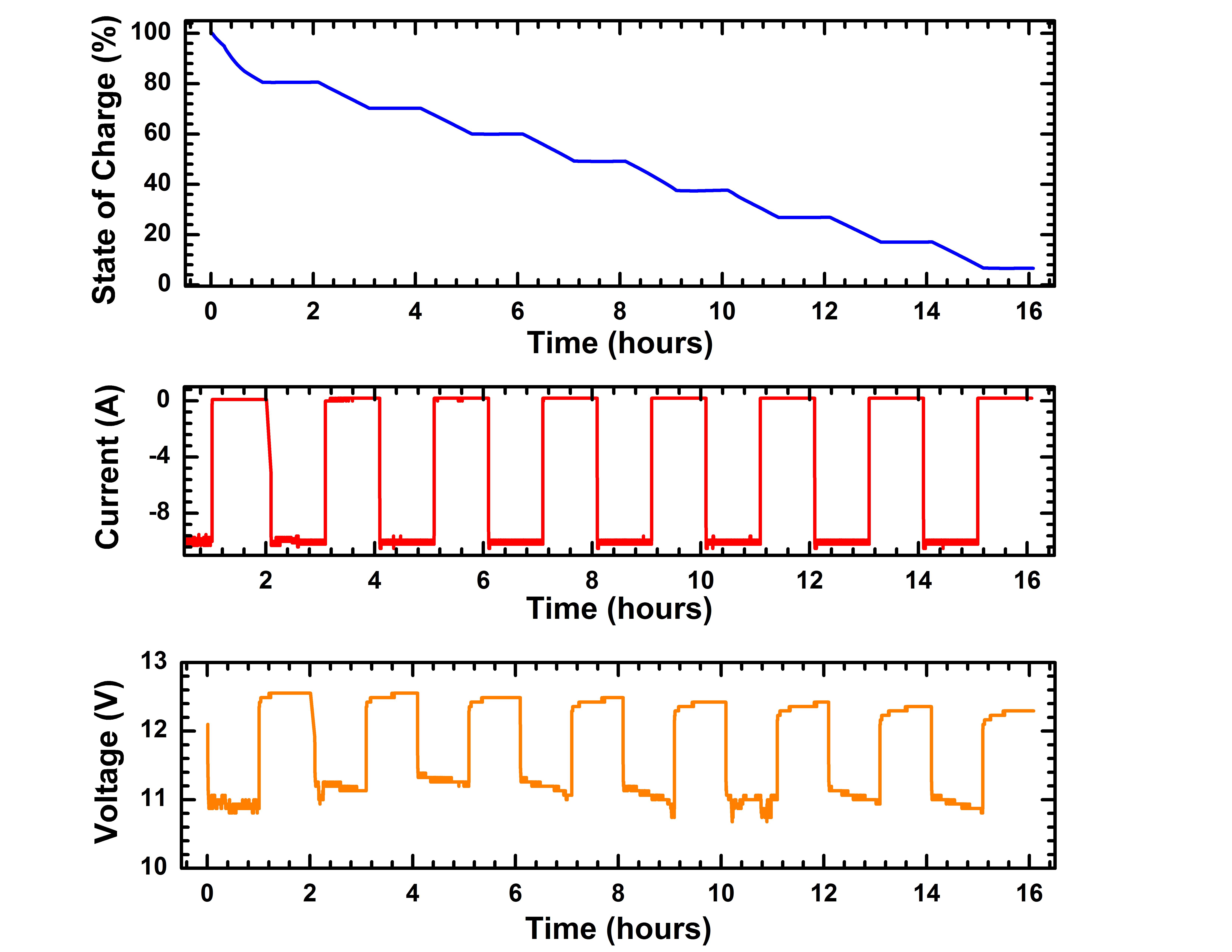}}
\caption{Estimated SoC using EKF}
\label{fig11}
\end{figure}

\begin{figure}[hbpt]
\captionsetup{justification=raggedright,singlelinecheck=false}
\centerline{\includegraphics[width = 0.8\columnwidth]{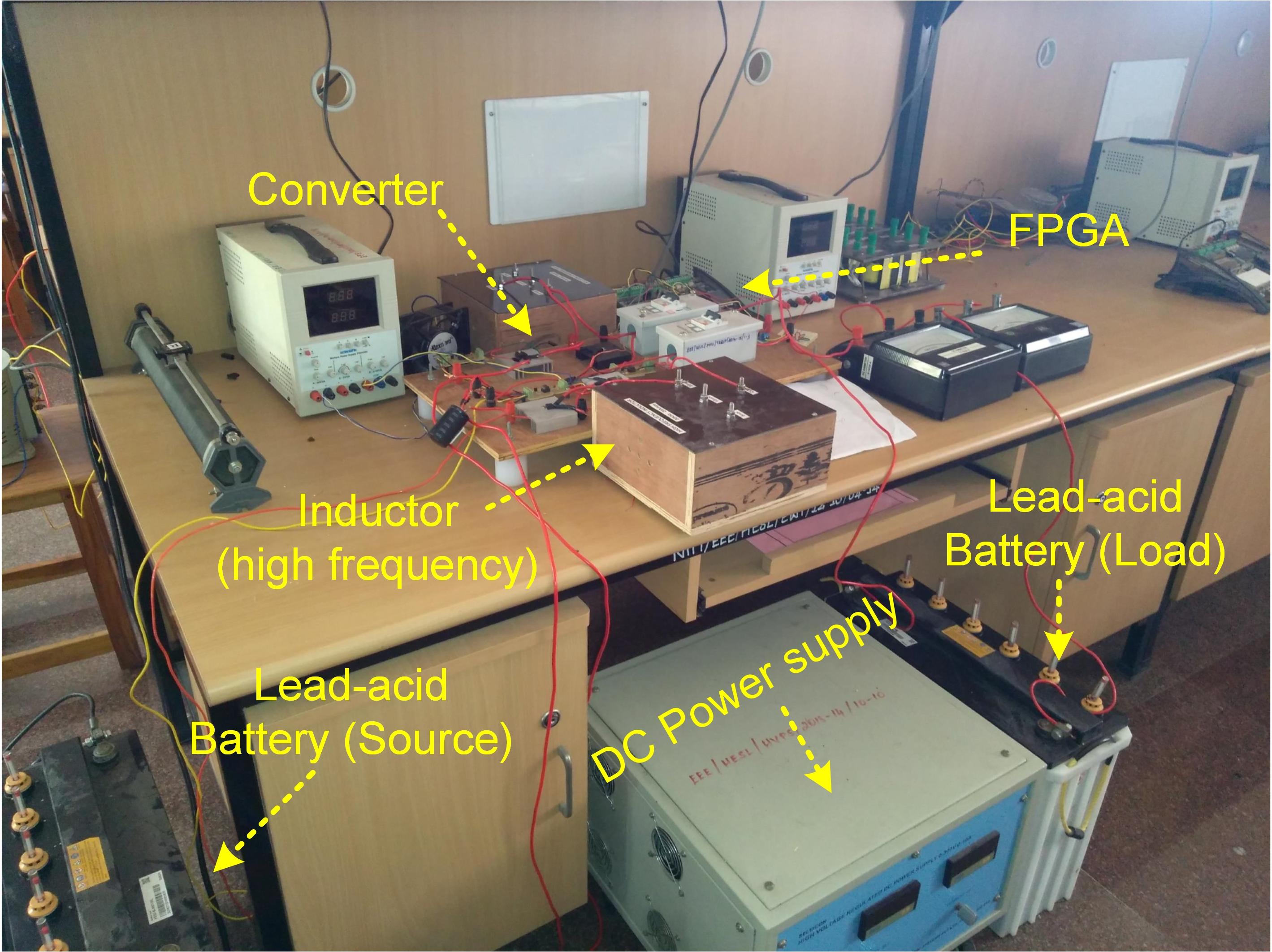}}
\caption{Experimental setup}
\label{fig12}
\end{figure}

An experiment is conducted using the developed controller after implementing algorithm in FPGA. During the experiment, a lead-acid battery is charged using DC supply. It is referred here as source lead-acid battery. The description of components used in this experiment is given in Table III. Upon charging, source battery is discharged across another lead-acid battery referred here as load lead-acid battery. The SoC of source battery, which is being discharged, is measured using the developed controller. A constant discharging scheme followed by rest period is adapted while dumping the charge of source battery to load battery (Fig.~\ref{fig11}). The experimental setup of developed controller, the variation in SoC, current and voltage of source battery, and a sample image of implemented touch-screen based GUI are depicted in Fig.~\ref{fig10}, Fig.~\ref{fig11}, and Fig.~\ref{fig12} respectively. Further, If $V \gets float~voltage$ during charging, then the constant voltage charging mode is performed for very small duration which brings $s(t)$ to $100\%$ from $\approx 94\%$ quickly and ensures safety.
\begin{table}[b]
\caption{Description of Components}
\begin{center}
\begin{tabular}{c|c|c}
\hline
\textbf{S.~No.} & \textbf{Item} & \textbf{Part Number} \\
\hline
1	&FPGA Board	&Altium Nanoboard 3000\\
2	&Battery	&Exide 6LMS 100L, 100Ah, C10\\
3	&Diode		&RHP30120\\
4	&IGBT		&FGA25N120ANTD\\
5	&Inductor	&20 kHz, 20 A, 20mH\\
6	&Optocoupler	&HP HCPL-3101\\
\hline
\end{tabular}
\label{tab3}
\end{center}
\end{table}

\section{Conclusion}
This paper has presented $2^{nd}$ order equivalent circuit model of a lead-acid battery. A converter with closed loop control using FPGA is designed which allows consumer to run tests in real time to obtain the internal parameters of battery. The constant current charging/discharging, an important feature of this controller, has become possible to implement by a control scheme using PID controller in FPGA. The touch-screen based GUI has provided ease of access and increased the adaptability of the controller as test bench for the battery. The proposed scheme can also be implemented in modern low-cost microcontrollers.

\bibliographystyle{IEEEtran}
\bibliography{IEEEabrv,Ref}
\end{document}